# Thermal cycling memory in phase separated manganites

Bernardo Sievers[a], Mariano Quintero[a,b], and Joaquín Sacanell[a,b,]*

[a] *Departamento de Física de la Materia Condensada, Gerencia de Investigación y Aplicaciones, Centro Atómico Constituyentes, CNEA, , Av. General Paz 1499, San Martín (1650), Provincia de Buenos Aires, Argentina.*
[b] *INN, CONICET-CNEA, Av. General Paz 1499, San Martín (1650), Provincia de Buenos Aires, Argentina.*

**Abstract**

We have studied the irreversibility of the magnetization induced by thermal cycles in $La_{0.5}Ca_{0.5}MnO_3$ manganites, which present a low temperature state characterized by the coexistence of phases. The effect is evidenced by a decrease of the magnetization after cycling the sample between 300 and 50 K. We developed a phenomenological model that allows us to correlate the value of the magnetization with the number of cycles performed. The experimental results show excellent agreement with our model, suggesting that this material could be used for the development of a device to monitor thermal changes. The effect of thermal cycling is towards an increase of the amount of the non ferromagnetic phase in the compounds and it might be directly related with the strain at the contact surface among the coexisting phases.

## 1. Introduction

Phase separated manganites [1], have been the focus of extensive research during the last 15 years due to the high interplay between their electric, magnetic and structural degrees of freedom. That interplay gives rise to a wide variety of physical properties with many potential technological applications, as for example the colossal magnetoresistance effect (CMR) [1], the nonvolatile memory effect (NVM) [2,3], the magnetocaloric effect (MCE) [4,5,6], a resistive switching effect (RS) induced by the application of an electric potential [7,8,9], etc. The phase separated state of manganites is characterized by the coexistence of different magnetic phases and is the key to understand those phenomena. Moreover, the presence of inhomogeneous states was observed to be important in many other strongly correlated electron systems than such as cobaltites and cuprates [10] among others and is being considered nowadays more a rule than an exception.

---
* Corresponding author. Tel.: +5411-6772-6757
*E-mail address:* sacanell@tandar.cnea.gov.ar

The dynamical evolution of the electric and magnetic properties is one of the most interesting phenomena of phase separated manganites. The effect is related with changes in the relative fraction of the coexisting phases [2,11,12,13,14,15,16]. It can be affected by composition[17], by the application of magnetic fields [2,3], electric fields [18,19,20] and temperature [21,22,23,24]. Although the effect is very interesting from the basic point of view, no clear applications have been yet proposed using the dynamical behavior of manganites nor a study has been reported regarding its influence on the above mentioned effects (CMR, NVM, MCE and RS).

In particular, a monotonous cumulative change of the electric and magnetic properties that can be induced by thermal cycles, is extremely relevant for several applications. For example, it can be used in a device to monitor thermal changes of solid state radiation detectors that must be maintained at low temperature [25]. Also, manganites with RS has been proposed as an alternative mechanism to develop resistive RAM memories for space and nuclear environments [26], regarding their resistance to radiation. Considering the periodic changes in temperature to which a satellite is subjected, a deep understanding of the thermal cycles effect and its influence on the physical properties is mandatory.

$La_{0.5}Ca_{0.5}MnO_3$ (LCMO), is a prototypical manganite exhibiting phase separation (PS) [27]. It is paramagnetic at room temperature, it presents a transition to a mainly ferromagnetic (FM) state on cooling below $T_C \sim 225K$ and then to a charge-ordered (CO) state at $T_{co} \sim 150K$, which is also antiferromagnetic (AF). However, the second transition is not completed and a fraction of the FM phase remains trapped in the CO host, giving rise to PS. The crystalline structure of both phases slightly differ,[28] causing the appearance of a stressed interface between them [21,29]. The presence of this kind of regions provokes irreversible changes each time the sample is cooled across the transition temperature [24,30], resembling the phenomenology of the martensitic transitions[31].

In this work we present a study of the irreversible changes in the magnetic properties of LCMO, which is a prototypical manganite with phase separation. The results will be discussed in the framework of the PS phenomena taking into account that it influences the CMR, NVM, RS and MCE effects.

## 2. Experimental

Polycrystalline samples of LCMO were synthesized by Liquid Mix. We followed the route detailed in [27] to obtain a sample with 950 nm average grain size. Chemical composition and crystalline structure were verified by EDS and X ray diffraction, respectively. Magnetization measurements were performed in a commercial vibrating sample magnetometer Versalab$^{TM}$ manufactured by Quantum Design.

## 3. Results

In figure 1 we present the Magnetization as a function of Temperature, for consecutive thermal cycles. We can see that Magnetization below 200 K monotonously decreases on each cycle. The low temperature magnetization of LCMO is a direct measure of the relative FM fraction [27,32]. Thus, our results indicate a cumulative reduction of the relative FM fraction at low temperatures.

The thermal cycling effect (TCE) can be in principle, used to monitor the number of low temperature thermal cycles performed to the sample simply by measuring the low temperature magnetization or the electrical resistivity. In figure 2(a) we show M(50K) as a function of the number of thermal cycles performed (*n*).

To describe the observed phenomena we use a simple phenomenological model. At low temperature, the sample is a mixture of FM and non-FM regions [33]. Such coexistence would lead to the appearance of interface effects due to the small structural differences between the coexisting phases that were previously mentioned, and thus the interface will be subjected to stress. The change of one phase into another in a given cycle is proportional to the total amount of phase capable to be converted until the system reaches the equilibrium state. The dependence of the FM fraction (*x*) as a function of *n* will follow equation (1) [30].

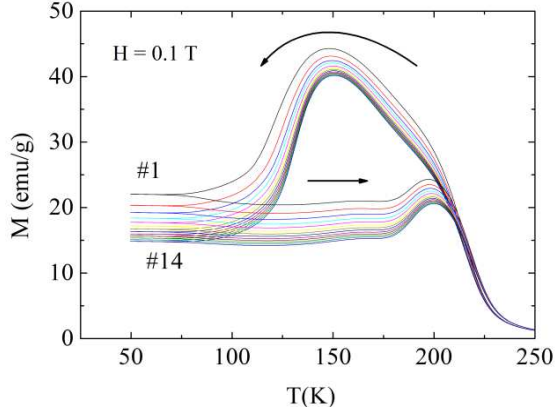

Figure 1: M vs T for a sequence of thermal cycles. H = 0.1 T.

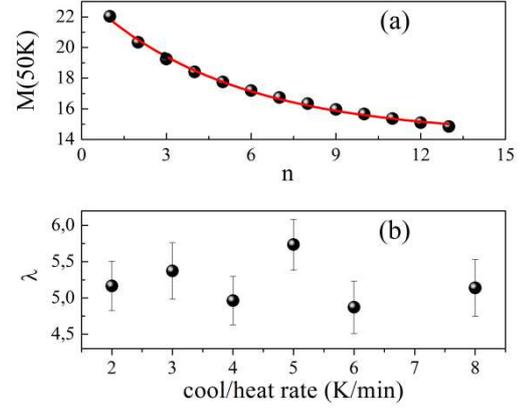

Figure 2: (a) M(50K) vs n. (b) Value of the experimental parameter λ, for different cooling and heating rates.

$$x(n) = x_0 e^{-n/\lambda} + x_{eq} \qquad (1)$$

In figure 2(a) we can see the well agreement between our experimental data for M(50K) and the proposed model, shown in solid line. Thus, λ plays a role of a "calibration" parameter which quantifies the relation between $x$ and $n$. We performed cycles varying the heating/cooling rate in order to analyze its influence on λ. In figure 2(b) we show the values of λ, obtained by fitting the data from sequences of thermal cycles performed at rates between 2 and 8 K/min with our model. We can see that the value of λ does not change within our experimental error, showing a value of $5.2 \pm 0.3$ as previously reported [30].

However, it is known that the electrical and magnetic properties of manganites with PS are subjected to a dynamical evolution that resembles a spin glass behavior [2,11,34]. Therefore, the question that arises is whether the TCE is related to that dynamic evolution or intrinsically related to a thermal induced effect that occurs each time the system undergoes a phase transition within the PS regime. It has been recently reported that the influence of dynamical features on the TCE depend on the T range in which the sample is retained (i.e. the dwell T) [30]. It has been shown that if the dwell T is above 100 K the value of λ is not affected, but if the dwell T is below 100 K, a qualitatively different behavior is observed. The change in $x$ in a cycle in which the sample is retained for several hours at dwell T below 100 K is larger than what it would be expected for a non-interrupted thermal cycle. However, in both cases (dwell T below and above 100 K), the initial value of λ is recovered once the sequence of thermal cycles starts again. Thus, if for some reason, the system is kept below 100K for a significant time, the system must be reseted by performing one or two extra cycles in order to recover the initial tendency.

## 4. Conclusions

In conclusion, we observed that thermal cycles induce changes in the low temperature magnetic properties of LCMO, which is a prototypical manganite exhibiting PS. The observed changes are related with variations in the relative content of the coexisting phases and are likely to occur due to the appearance of defects in the interfaces between the FM and CO phases. This means that the TCE must be carefully considered to avoid undesired shifts in novel applications of manganites (such as NVM, RS, MCE) where the temperature could be not completely controlled. On the other hand, as a positive aspect, since the effect is independent on the heating and cooling rates (for rates between 2 and 8 K/min), we propose that it can be used in a device to monitor thermal changes. We presented a phenomenological model in order to quantify the relation among the magnetization at low temperature and the number of thermal cycles experienced. Our model efficiently describes the effect even if the sample has been maintained for comparatively long periods of time at T > 100 K, beginning to fail if is kept at T < 100 K In both situations, the effect is recovered after two subsequent thermal cycles.

As a final remark, considering that the transition temperatures of manganites with PS can be tuned by chemical composition, a broad spectrum opens for the development of devices that can be used in different temperature ranges.

We are now performing a detailed study on the influence of the isothermal relaxations in order to consider them to develop a prototypical device.